\RequirePackage[2020-02-02]{latexrelease}
\documentclass[aps,prd,groupedaddress,graphicx,nofootinbib]{revtex4}
\usepackage{graphics,graphicx,color,epsf}
\usepackage{subfigure}
\usepackage[scanall]{psfrag}  
\usepackage{tikz}

\begin{document}

\title{On the nature of cosmic strings in the brane world}

\author{Chia-Min Lin}

\affiliation{Fundamental General Education Center, National Chin-Yi University of Technology, Taichung 41170, Taiwan}



\begin{abstract}

We investigate a static, cylindrically symmetric cosmic string on the brane without a perturbative approximation. We find there could be a (large) enhancement of the (effective) string tension when the energy density at the center of the string is (much) larger than twice the brane tension. We also point out a new way to evade the cosmic string problem when the energy density at the center of the string approaches twice the brane tension. These findings could have experimental and theoretical implications for searching for cosmic strings on the brane, in particular for cosmic strings generated after inflation (such as D-term inflation) on the brane. 

\end{abstract}
\maketitle
\large
\baselineskip 18pt
\section{Introduction}

The production of cosmic strings is quite generic in the framework of Grand Unified Theories (GUT) \cite{Jeannerot:2003qv}.
A seminal work in studying the gravitational effects of cosmic strings by a linear approximation to general relativity is given in \cite{Vilenkin:1981zs}. It is found that the spacetime is conical outside a static, cylindrically symmetric cosmic string. The results are extended to the exact spacetime metric in \cite{Hiscock:1985uc}. 
A study of cosmic string on the brane by linear approximation is given in \cite{Davis:2000uf}.
The method of finding the spacetime metric without a linear approximation is applied to cosmic strings on the brane in \cite{Abdalla:2015vna}. As we would see in the following sections, particular assumptions have to be made in order to obtain simple results that can be compared with previous results in general relativity. In this work, we extend the results of \cite{Abdalla:2015vna} and derive new formulas. In particular, we consider the role played by the four-dimensional cosmological constant. These results are general and could be applied to a wide range of cosmic strings on the brane. In order to have a concrete example, we use a notation with an eye toward its potential application to the cosmic strings generated after D-term inflation on the brane \cite{Lin:2022gbl}.

Cosmic strings produce stochastic gravitational waves \cite{Hindmarsh:1994re} that can be constrained via experiments such as European Pulsar Timing Array (EPTA) \cite{vanHaasteren:2011ni}, NANOGrav Collaboration \cite{NANOGrav:2020bcs, Ellis:2020ena, Blasi:2020mfx}, LIGO-Virgo \cite{LIGOScientific:2017ikf}, and Laser Interferometer Space Antenna (LISA) \cite{Auclair:2019wcv}. These observations constrain the cosmic string tension $\mu$\footnote{Usually expressed as $G\mu$. Here $G$ is Newton's constant and $\mu$ is mass per unit length.}. We would like to investigate how the string tension is modified in the framework of a braneworld.

As an example, cosmic strings on the brane can be generated if we consider a D-term inflation \cite{Binetruy:1996xj, Halyo:1996pp} on the brane.
The potential energy density of D-term inflation is
\begin{equation}
V \simeq V_0=\frac{g^2\xi^2}{2},
\label{eq1}
\end{equation}
where $\xi$ is the Fayet-Iliopoulos term and $g$ is the $U(1)_{FI}$ gauge coupling. After inflation, the symmetry-breaking field developed a vacuum expectation value of $\sqrt{2\xi}$ and form a network of cosmic strings. Conventional D-term inflation predicts $\sqrt{2\xi}$ to be around the scale of grand unified theories (GUT) which generates cosmic strings with too large tension to be compatible with current experimental observations. This is referred to as a cosmic string problem. We will discuss the implication of our finding to this problem.

\section{brane world}
The brane world models of Randall and Sundrum \cite{Randall:1999ee, Randall:1999vf} have stimulated vast interest in higher-dimensional theories.
If our four-dimensional world is a 3-brane embedded in a higher-dimensional bulk, the four-dimensional Einstein equations induced on the brane are given by the covariant approach as \cite{Shiromizu:1999wj} (see \cite{Maartens:2010ar} for a review and more references therein.)
\begin{equation}
G_{\mu \nu}=-\Lambda_4 g_{\mu\nu}+\left( \frac{8\pi}{M_4^2} \right) T_{\mu\nu}+\left( \frac{8\pi}{ M_5^3} \right)^2 \Pi_{\mu\nu}-E_{\mu\nu}.
\label{branee}
\end{equation}
In the above equation, $T_{\mu\nu}$ is the energy-momentum tensor of matter on the brane, and
\begin{equation}
\Pi_{\mu\nu}=-\frac{1}{4} T^\sigma_\mu T_{\nu \sigma}+\frac{1}{12}TT_{\mu\nu}+\frac{1}{8}g_{\mu\nu}T_{\alpha\beta}T^{\alpha\beta}-\frac{1}{24}g_{\mu\nu}T^2,
\label{qqq}
\end{equation}
which is quadratic in $T_{\mu\nu}$. The last term $E_{\mu\nu}$ is from the projection of the bulk Weyl curvature on the brane, which can be expressed as a Weyl fluid \cite{Maartens:2010ar},
\begin{equation}
-E_{\mu\nu}=8 \pi G \left[ U \left( u_\mu u_\nu -\frac{1}{3}h_{\mu\nu} \right)+P_{\mu\nu}+Q_\mu u_\nu+Q_\nu u_\mu \right],
\label{eqe}
\end{equation}
where $U$ is the energy density of dark radiation\footnote{It incorporates the spin-0 mode of the 5D graviton.}, $P_{\mu\nu}$ is the anisotropic pressure, and $Q_\mu$ is the momentum density.
The metric tensor is decomposed by a 4-velocity $u_\mu$ as $g_{\mu\nu}=h_{\mu\nu}+u_\mu u_\nu$.
The four-dimensional cosmological constant $\Lambda_4$ is determined by the five-dimensional bulk cosmological constant $\Lambda_5$ and the brane tension $\Lambda$ as 
\begin{equation}
\Lambda_4=\frac{4\pi}{M_5^3}\left(\Lambda_5+\frac{4\pi}{3M_5^3} \Lambda^2 \right),
\end{equation}
which can be set to $\Lambda_4=0$ by asuming a suitable $\Lambda_5$. 
The brane tension $\Lambda$ provides a relation between the four-dimensional Planck scale $M_4=\sqrt{8 \pi}M_P$ and five-dimensional Planck scale $M_5$ via
\begin{equation}
M_4=\sqrt{\frac{3}{4\pi}}\left( \frac{M_5^2}{\sqrt{\Lambda}} \right)M_5.
\label{pl}
\end{equation}
By using Eq.~(\ref{pl}) and $1/M_4^2=G$, Eq.~(\ref{branee}) can be expressed as
\begin{equation}
G_{\mu \nu}=-\Lambda_4 g_{\mu\nu}+8\pi G T_{\mu\nu}+\frac{48\pi G}{\Lambda} \Pi_{\mu\nu}-E_{\mu\nu} \equiv 8\pi G \overline{T}_{\mu\nu}.
\label{branee2}
\end{equation}
Here we defined an effective energy momentum tensor $\overline{T}_{\mu\nu}$.
If we assume that there is no energy-momentum exchange between the bulk and the brane, the energy conservation is satisfied both for the matter on the brane $T_{\mu\nu}$ and for $\overline{T}_{\mu\nu}$, therefore 
\begin{eqnarray}
\nabla^\mu T_{\mu\nu}&=&0,   \label{tcon} \\ 
\nabla^\mu \left( \frac{48\pi G}{\Lambda}\Pi_{\mu\nu}-E_{\mu\nu} \right)&=&0.
\end{eqnarray}

\section{cosmic string on the brane}

 For simplicity, we consider a straight cosmic string produced after D-term inflation on the brane in cylindrical coordinates $\{t, \rho, \phi, z\}$. 
The energy-momentum tensor of a cosmic string is represented by \cite{Vilenkin:1981zs}
\begin{equation}
T^\nu_\mu=-V_0\mbox{ diag}(1,0,0,1).
\label{emt}
\end{equation}
We assume for $\rho < \rho_0$, the energy density is constant $V_0=g^2 \xi^2/2$, and for $\rho > \rho_0$ it is zero. 
By using the vacuum expectation value $\sqrt{2\xi}$ of the symmetry breaking field, the value of $\rho_0$ can be obtained by the balance between ``spatial variation'' term and potential term from Eq.~(\ref{eq1}) as
\begin{equation}
\left( \frac{\sqrt{2\xi}}{\rho_0} \right)^2 \sim \frac{g^2\xi^2}{2}.
\end{equation}
This implies\footnote{In general, one may assume the vacuum expectation value of the symmetry-breaking field to be $\eta$ with energy density $\eta^4$ at the center of the cosmic string and estimates $\rho_0=1/\eta$. Our calculation applies to general cases.} 
\begin{equation}
\rho_0=\frac{2}{g\sqrt{\xi}}.
\end{equation}
By using the cylinderical symmetry, the line element is
\begin{equation}
ds^2=-e^{2\alpha(\rho)}dt^2+e^{2\beta(\rho)}d\rho^2+e^{2\gamma(\rho)}d\phi^2+e^{2\delta(\rho)}dz^2.
\label{metric}
\end{equation}
From Eqs.~(\ref{qqq}) and (\ref{emt}), the only non-vanishing components of $T^\nu_\mu$ are
\begin{equation}
\Pi^1_1=\Pi^2_2=\frac{1}{12}V_0^2.
\end{equation}
In order to have a static and cylindrically symmetric metric, the momentum density is assumed to be $Q_\mu=0$, and Eq.~(\ref{eqe}) can be arranged as
\begin{equation}
-E^\nu_\mu=8 \pi G \mbox{ diag}\left[ -U,P_1+\frac{1}{3}U, P_2+\frac{1}{3}U,P_3+\frac{1}{3}U\right],
\end{equation}
where $U$, $P_1$, $P_2$, and $P_3$ are functions of $\rho$. Because $E^\nu_\mu$ is the projection of the Weyl curvature tensor, it is traceless. This implies
\begin{equation}
P_1+P_2+P_3=0.
\label{ppp}
\end{equation}
From Eq.~(\ref{metric}), we can calculate $G_{\mu\nu}$. Some details are given in the Appendix section. 
By using (\ref{branee}), when $\rho < \rho_0$,
\begin{eqnarray}
e^{-2\beta}(\gamma^{\prime 2}+\alpha^{\prime 2}-\beta^\prime \gamma^\prime+\beta^\prime \alpha^\prime-\gamma^\prime\alpha^\prime +\gamma^{\prime\prime}-\alpha^{\prime\prime})&=&-8\pi G V_0 -8\pi G U-\Lambda_4,\\
-e^{-2\beta}\alpha^{\prime 2}&=&8\pi G\frac{V_0^2}{2\Lambda}+\frac{8\pi G}{3}U+8\pi G P_1-\Lambda_4,     \label{e16}\\
e^{-2\beta}\alpha^{\prime 2}&=&8\pi G\frac{V_0^2}{2\Lambda}+\frac{8\pi G}{3}U+8\pi G P_2-\Lambda_4,    \label{e17} \\ 
e^{-2\beta}(\gamma^{\prime 2}+\alpha^{\prime 2}-\beta^\prime \gamma^\prime-\beta^\prime \alpha^\prime+\gamma^\prime\alpha^\prime +\gamma^{\prime\prime}+\alpha^{\prime\prime})&=&-8\pi G V_0 +\frac{8\pi G}{3}U+8\pi G P_3-\Lambda_4,
\end{eqnarray}
From Eq.~(\ref{tcon}), we have
\begin{equation}
\nabla^\mu T_{\mu 1}=V_0(\alpha^\prime +\delta^\prime)=0,
\end{equation}
which gives $\alpha+\delta=\mbox{constant}$. The constant can be chosen to be zero by a coordinate transformation. Therefore  
\begin{equation}
\delta=-\alpha.
\end{equation}
We wish the right-hand sides of Eqs.~(\ref{e16}) and (\ref{e17}) to be zero in order to obtain a form of $\overline{T}^\nu_\mu$ similar to Eq.~(\ref{emt}). Therefore $\alpha=\mbox{constant}$, which again can be chosen to be zero. Actually, having $\alpha=\delta=0$ here is equivalent to imposing Lorentz invariance in the $z$ direction. Finally, there is still a degree of freedom of $\beta$ to be chosen to $\beta=\delta=0$ through a coordinate (gauge) transformation \cite{Chandrasekhar:1972zz}. Therefore
\begin{eqnarray}
\gamma^{\prime 2}+\gamma^{\prime\prime}&=&-8\pi G V_0 -8\pi G U-\Lambda_4,  \label{e22} \\
0&=&8\pi G\frac{V_0^2}{2\Lambda}+\frac{8\pi G}{3}U+8\pi G P_1-\Lambda_4,    \\
0&=&8\pi G\frac{V_0^2}{2\Lambda}+\frac{8\pi G}{3}U+8\pi G P_2-\Lambda_4,    \\ 
\gamma^{\prime 2}+\gamma^{\prime\prime}&=&-8\pi G V_0 +\frac{8\pi G}{3}U+8\pi G P_3-\Lambda_4. \label{e24}
\end{eqnarray}

\subsection{Assuming $\Lambda_4=0$}

\begin{figure}[t]
  \centering
\includegraphics[width=0.6\textwidth]{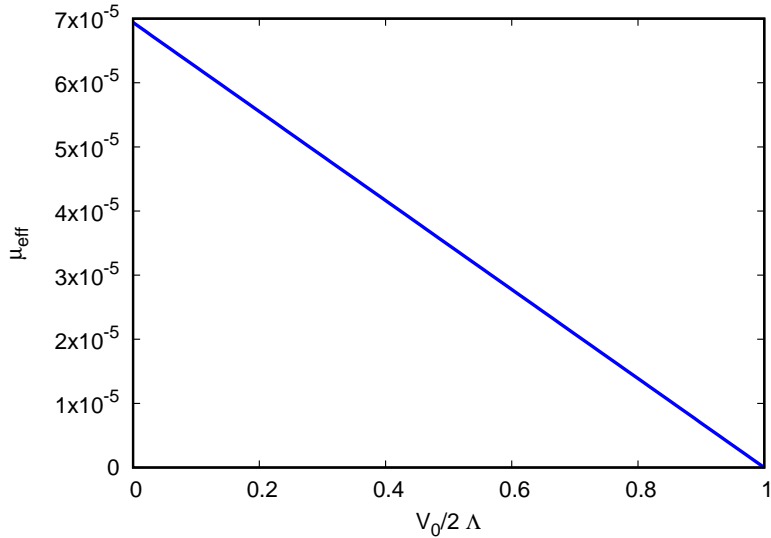}
  \caption{$\mu_{\mathrm{eff}}$ as a function of time $V_0/ 2\Lambda$ for $\Lambda_4=0$ and $V_0/ 2\Lambda \leq  1$. Here we have setted $8\pi G=1$ and use a typical value $\xi=1.1 \times 10^{-5}$.}
  \label{fig1}
\end{figure}

Firstly, let us set $\Lambda_4=0$ as was done in \cite{Abdalla:2015vna}. By using Eq.~(\ref{ppp}) and the above equations, one obtains
\begin{eqnarray}
U&=&-\frac{V_0^2}{2\Lambda}   \\
P_3&=&-2P_1=-2P_2=-\frac{4}{3}U.
\end{eqnarray}
Note that the required energy density of dark radiation is negative and it is only confined within the cosmic string at $\rho < \rho_0$.
Substituting these into Eq.~(\ref{e22}) (or (\ref{e24})), we have
\begin{equation}
\gamma^{\prime 2}+\gamma^{\prime\prime}=-8\pi G V_0\left( 1-\frac{V_0}{2\Lambda} \right). 
\label{eq28}
\end{equation}
The solution of $\gamma$ depends on whether $V_0<2\Lambda$ or $V_0>2\Lambda$. In the case $V_0<2\Lambda$ the solution is
\begin{equation}
\gamma=\ln \left[ \rho_\ast \sin \left( \frac{\rho}{\rho_\ast } \right) \right],
\label{gamma1}
\end{equation}
where
\begin{equation}
\rho_\ast=\frac{1}{\sqrt{8\pi G V_0\left( 1-\frac{V_0}{2\Lambda} \right)}}.
\end{equation}
The integration constants are fixed by requiring the metric on the axis to be flat without cone singularity.
Compare Eq.~(\ref{emt}) with Eq.~(\ref{eq28}), the effective energy density of the cosmic string is
\begin{equation}
-\overline T^0_0=\frac{G^0_0}{-8\pi G}=V_0\left( 1-\frac{V_0}{2\Lambda} \right).
\end{equation}
The effective cosmic string tension is given by\footnote{Our result is different from that of \cite{Abdalla:2015vna}.}
\begin{eqnarray}
\mu_{\mathrm{eff}}&=&\int^{2\pi}_0\int^{\rho_0}_0 V_0\left( 1- \frac{V_0}{2\Lambda} \right)\rho_\ast \sin  \left( \frac{\rho}{\rho_\ast } \right) d\phi d\rho   \\
  &=&\frac{1}{4G}  \left[ 1- \cos \left(  \sqrt{2\xi}\sqrt{8\pi G \left( 1-\frac{V_0}{2\Lambda} \right)}      \right) \right]. \label{mu1}
\end{eqnarray}
This is plotted in Fig. \ref{fig1}.
By expanding $\cos x \sim 1-x^2/2$ we have
\begin{equation}
\mu_{\mathrm{eff}}=2\pi \xi \left( 1-\frac{V_0}{2\Lambda} \right),
\label{f1}
\end{equation}
which reproduces the standard result from general relativity when $V_0 \ll 2\Lambda$. 
The result of $\gamma$ in Eq.~(\ref{gamma1}) can be connected with a solution $\gamma=\ln(ar)$ outside the cosmic string with a radial distance $r$ by matching the junction conditions $\gamma(\rho_0)=\gamma(r_0)$ and $\gamma^\prime(\rho_0)=\gamma^\prime(r_0)$ at $\rho_0=r_0$ to obtain\footnote{The junction conditions are that the (induced) metric and the extrinsic curvature be the same on both sides of the hypersurface \cite{Poisson:2009pwt}. On the hypersurface of $r_0=\rho_0$, the unit normal vector is $n^\mu=(0,0,e^{-\gamma},0)$, and the non-vanishing extrinsic curvature is $K_{12}=-2\gamma^\prime e^{-\gamma}$.}
\begin{equation}
a= \cos \left( \sqrt{2\xi} \sqrt{8\pi G \left( 1-\frac{V_0}{2\Lambda} \right)} \right)=1-4G\mu_{\mathrm{eff}},
\end{equation} 
where the second equality is obtained from Eq.~(\ref{mu1}). With $\gamma=\ln (ar)$ (and $\alpha=\beta=\delta=0$), the metric outside the cosmic string is\footnote{In \cite{Heydari-Fard:2013eoa}, this form of metric is assumed without a calculation of string tension.}
\begin{equation}
ds^2=-dt^2+dr^2+(1-4G \mu_{\mathrm{eff}})^2 r^2 d\phi^2.
\label{ex}
\end{equation}
When $\phi$ changes by $2\pi$, the effective angular coordinate $\bar{\phi}=(1-4G \mu_{\mathrm{eff}})\phi$ changes by $2\pi-\Delta$.
The deficit angle $\Delta$ is
\begin{equation}
\Delta=8\pi G  \mu_{\mathrm{eff}} = 2\pi(1-a).
\end{equation}

\begin{figure}[t]
  \centering
\includegraphics[width=0.6\textwidth]{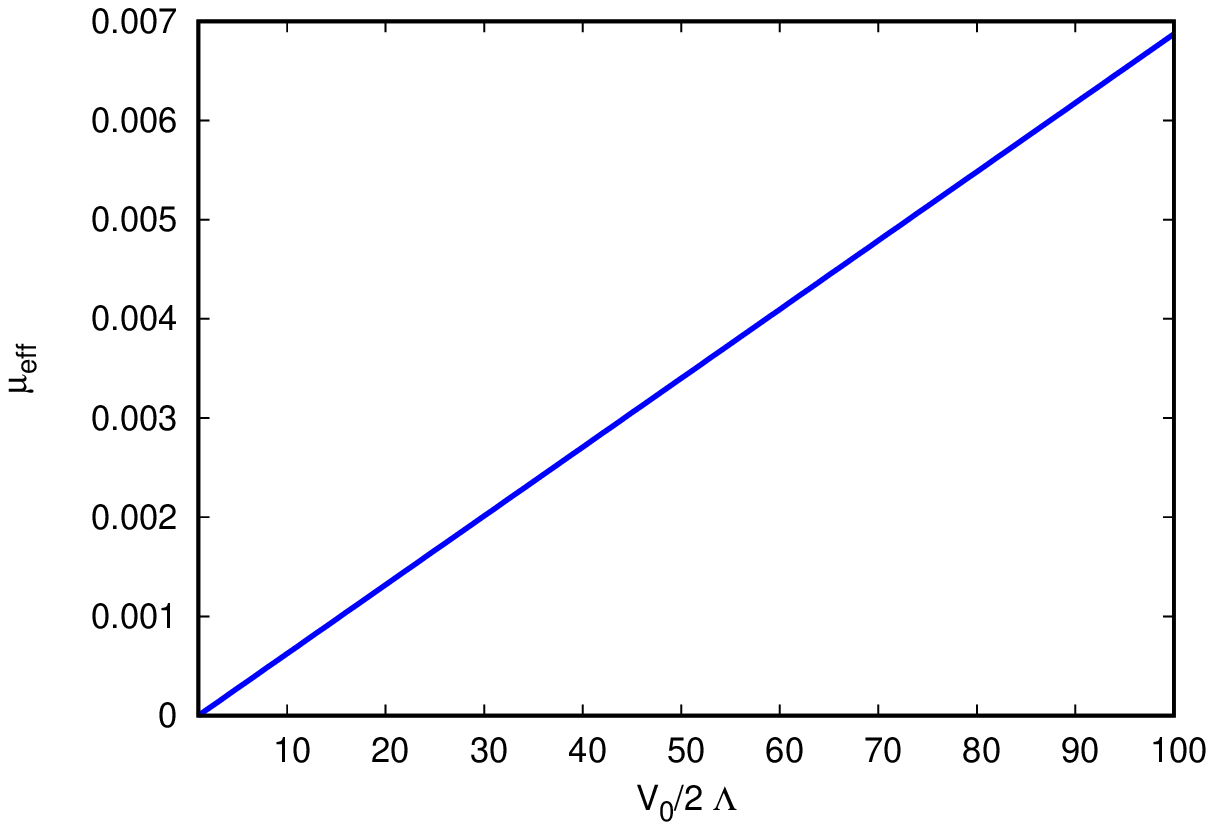}
  \caption{$\mu_{\mathrm{eff}}$ as a function of time $V_0/ 2\Lambda$ for $\Lambda_4=0$ and $V_0/ 2\Lambda \geq 1$. Here we have setted $8\pi G=1$ and use a typical value $\xi=1.1 \times 10^{-5}$.}
  \label{fig2}
\end{figure}

On the other hand, we are interested in having $V_0>2\Lambda$. In this case, the solution of Eq.~(\ref{eq28}) is given by
\begin{equation}
\gamma(\rho)=\ln \left[ \rho_\ast \sinh \left( \frac{\rho}{\rho_\ast } \right) \right],
\end{equation}
where
\begin{equation}
\rho_\ast=\frac{1}{\sqrt{8\pi G V_0\left( \frac{V_0}{2\Lambda} -1 \right)}}.
\end{equation}
The effective cosmic string tension is 
\begin{eqnarray}
\mu_{\mathrm{eff}}&=&\int^{2\pi}_0\int^{\rho_0}_0 V_0\left( \frac{V_0}{2\Lambda}-1 \right)\rho_\ast \sinh  \left( \frac{\rho}{\rho_\ast } \right) d\phi d\rho   \\
  &=&\frac{1}{4G} \left[ \cosh \left(  \sqrt{2\xi}\sqrt{8\pi G \left( \frac{V_0}{2\Lambda}-1 \right)}      \right)-1 \right].
\end{eqnarray}
This is plotted in Fig. \ref{fig2}.
By expanding $\cosh x \sim 1+x^2/2$ we have
\begin{equation}
\mu_{\mathrm{eff}}=2\pi \xi  \left( \frac{V_0}{2\Lambda}-1 \right).
\label{mu2}
\end{equation}
If $V_0 \gg 2\Lambda$, there is a big enhancement of the cosmic string tension compared with the case of $V_0 \ll 2\Lambda$.
In this case, we have
\begin{equation}
a= \cosh \left( \sqrt{2\xi} \sqrt{8\pi G \left( \frac{V_0}{2\Lambda} -1\right)} \right)>1,
\end{equation} 
which implies a negative deficit angle $\Delta=2\pi(1-a)$. In this case, there is no duplicated images from gravitational lensing.

One interesting case happens when $V_0=2\Lambda$. From Eq.~(\ref{mu2}) (or Eq.~(\ref{f1})), we have $\mu_{\mathrm{eff}}=0$ and $\Delta=0$. This is due to the assumption that dark radiation with negative energy density exists inside the cosmic string. If this is true, it could be a method to solve the cosmic string problem in D-term inflation. In order to evade the cosmic string problem, we only need $V_0$ to be close enough to $2\Lambda$ to obtain a string tension within experimental bounds.

\subsection{Assuming $E_{\mu\nu}=0$}
\begin{figure}[t]
  \centering
\includegraphics[width=0.6\textwidth]{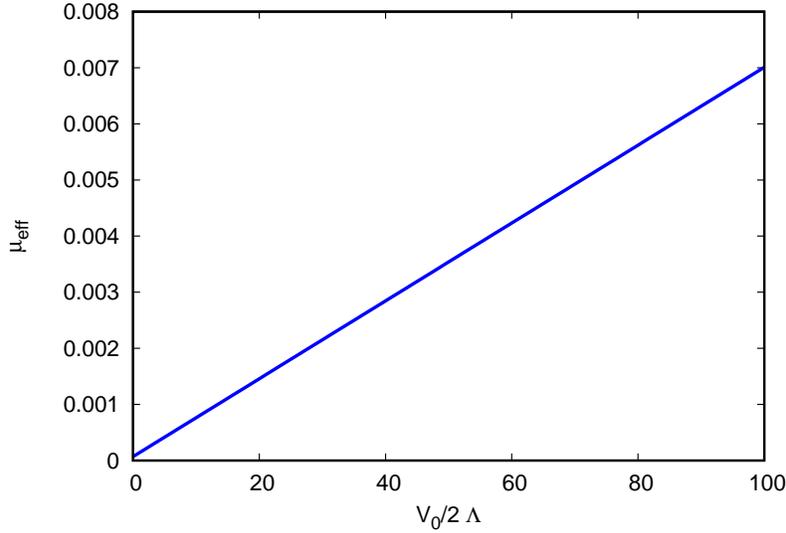}
  \caption{$\mu_{\mathrm{eff}}$ as a function of time $V_0/ 2\Lambda$ for $E_{\mu\nu}=0$. Here we have setted $8\pi G=1$ and use a typical value $\xi=1.1 \times 10^{-5}$.}
  \label{fig3}
\end{figure}

In the above discussion, we have assumed $\Lambda_4=0$, now we consider the case $E_{\mu\nu}=0$. namely $U=P_1=P_2=P_3=0$. We have
\begin{eqnarray}
\gamma^{\prime 2}+\gamma^{\prime\prime}&=&-8\pi G V_0 -\Lambda_4,  \label{e40} \\
0&=&8\pi G\frac{V_0^2}{2\Lambda}-\Lambda_4.   \label{e41}  
\end{eqnarray}
From Eq.~(\ref{e41}), we obtain
\begin{equation}
\Lambda_4=8\pi G \frac{V_0^2}{2\Lambda}.
\end{equation}
Substitute this into Eq.~(\ref{e40}) gives
\begin{equation}
\gamma^{\prime 2}+\gamma^{\prime\prime}=-8\pi G V_0 -8\pi G \frac{V_0^2}{2\Lambda}.
\label{40}
\end{equation}
The solution is
\begin{equation}
\gamma=\ln \left[ \rho_\ast \sin \left( \frac{\rho}{\rho_\ast } \right) \right],
\end{equation}
where
\begin{equation}
\rho_\ast=\frac{1}{\sqrt{8\pi G V_0\left( 1-\frac{V_0}{2\Lambda} \right)}}.
\end{equation}
Compare Eq.~(\ref{emt}) with Eq.~(\ref{40}), the effective energy density of the cosmic string is
\begin{equation}
\frac{G^0_0}{-8\pi G}=V_0\left( 1+\frac{V_0}{2\Lambda} \right).
\end{equation}
The effective cosmic string tension is given by
\begin{eqnarray}
\mu_{\mathrm{eff}}&=&\int^{2\pi}_0\int^{\rho_0}_0 V_0\left( 1+ \frac{V_0}{2\Lambda} \right)\rho_\ast \sin  \left( \frac{\rho}{\rho_\ast } \right) d\phi d\rho   \\
  &=&\frac{1}{4G}  \left[ 1- \cos \left(  \sqrt{2\xi}\sqrt{8\pi G \left( 1+\frac{V_0}{2\Lambda} \right)}      \right) \right].
\end{eqnarray}
This is plotted in Fig. \ref{fig3}.
By expanding $\cos x \sim 1-x^2/2$ we have
\begin{equation}
\mu_{\mathrm{eff}}=2\pi \xi  \left( 1+\frac{V_0}{2\Lambda} \right)   
\label{f3}
\end{equation}
In this case, we have
\begin{equation}
a= \cos \left( \sqrt{2\xi} \sqrt{8\pi G \left( 1+\frac{V_0}{2\Lambda} \right)} \right).
\end{equation} 
This result can be applied for both $V_0>2\Lambda$ and $V_0< 2\Lambda$. The deficit angle $\Delta=2\pi(1-a)$ would always be positive in this case.

Although we consider two cases $\Lambda_4=0$ or $E_{\mu\nu}=0$, in general, they can both be non-zero which gives 
\begin{equation}
U=\frac{\Lambda_4}{8\pi G}-\frac{V_0^2}{2\Lambda}.
\end{equation}
This provides a chance to have a positive energy density for dark radiation.
\section{Conclusion}
\label{con}

We have calculated three different forms of $\mu_{\mathrm{eff}}$. The main findings of this work are Eqs.~(\ref{f1}), (\ref{mu2}), and (\ref{f3}). Experiments searching for cosmic strings usually provide constraints of the dimensionless quantity $G\mu$. In the context of cosmic strings on the brane, the constraints have to be imposed on $G\mu_{\mathrm{eff}}$ instead of $G\mu$. This would have significant implications for the search for cosmic strings if we live in a brane world. One interesting observation is that when $V_0=2\Lambda$, we have $\mu_{\mathrm{eff}}=0$ from Eqs.~(\ref{f1}) and (\ref{mu2}). This provides a novel method to deal with the cosmic string problem. Another effect is when $V_0 \gg 2\Lambda$, there is a big enhancement of $\mu_{\mathrm{eff}}$ from Eq.~(\ref{mu2}) and (\ref{f3}). This potentially could make the cosmic string problem more severe.


\appendix

\section{Useful formulas}
The non-zero Christoffel symbols are 
\begin{eqnarray}
\Gamma^1_{00}&=&\alpha^\prime e^{2(\alpha-\beta)}\\
\Gamma^0_{10}&=&\alpha^\prime \\
\Gamma^1_{11}&=&\beta^\prime  \\
\Gamma^2_{21}&=&\gamma^\prime  \\
\Gamma^1_{22}&=&-\gamma^\prime e^{2(\gamma-\beta)}  \\
\Gamma^3_{31}&=&\delta^\prime  \\
\Gamma^1_{33}&=&-\delta^\prime e^{2(\delta-\beta)}  \\
\end{eqnarray}
The Ricci tensors are
\begin{eqnarray}
R_{00}&=&(\alpha^{\prime\prime}+\alpha^{\prime 2}-\alpha^\prime \beta^\prime +\alpha^\prime \gamma^\prime + \alpha^\prime \delta^\prime)e^{2(\alpha-\beta)}   \\
R_{11}&=&\alpha^\prime \beta^\prime + \gamma^\prime \beta^\prime + \delta^\prime \beta^\prime - \gamma^{\prime\prime} -\delta^{\prime\prime}-\alpha^{\prime 2}-\gamma^{\prime 2}-\delta^{\prime 2}-\alpha^{\prime\prime}   \\
R_{22}&=&(-\gamma^{\prime\prime} + \gamma^\prime \beta^\prime-\alpha^\prime \gamma^\prime - \delta^\prime \gamma^\prime -\gamma^{\prime 2})e^{2(\gamma-\beta)}   \\
R_{33}&=&(-\delta^{\prime\prime}+\delta^\prime \beta^\prime - \alpha^\prime \delta^\prime - \gamma^\prime \delta^\prime -\delta^{\prime 2})e^{2(\delta-\beta)}
\end{eqnarray} 
The Ricci scalar is
\begin{equation}
R=(-2 \alpha^{\prime\prime}-2\alpha^{\prime 2}+2\alpha^\prime \beta^\prime -2 \alpha^\prime \gamma^\prime -2\alpha^\prime \delta^\prime +2\gamma^\prime \beta^\prime +2 \delta^\prime \beta^\prime -2\gamma^{\prime\prime}-2\delta^{\prime\prime}-2\gamma^{\prime 2}-2\delta^{\prime 2}-2\delta^\prime \gamma^\prime)e^{-2\beta}
\end{equation}
The Einstein tensors are
\begin{eqnarray}
G^0_0&=&(\gamma^{\prime 2}+\alpha^{\prime 2}-\beta^\prime \gamma^\prime + \beta^\prime \alpha^\prime -\gamma^\prime \alpha^\prime + \gamma^{\prime\prime}-\alpha^{\prime\prime})e^{-2\beta}\\
G^1_1&=&-\alpha^{\prime 2}e^{-2\beta} \\
G^2_2&=&\alpha^{\prime 2}e^{-2\beta} \\
G^3_3&=&(\alpha^{\prime\prime}+\alpha^{\prime 2}-\alpha^\prime \beta^\prime + \alpha^\prime \gamma^\prime -\gamma^\prime \beta^\prime +\gamma^{\prime\prime}+\gamma^{\prime 2})e^{2(\delta-\beta)}
\end{eqnarray}

\acknowledgments
This work is supported by the National Science and Technology Council (NSTC) of Taiwan under Grant No. NSTC 111-2112-M-167-002.

\end{document}